\documentclass[]{spie}  

\usepackage[]{graphicx}

\title{Scale-free networks in complex systems} 

\author{M. Bartolozzi\supit{a,b}, D.B. Leinweber\supit{a,b}, T. Surungan\supit{a,b,c}, A.W. Thomas\supit{a,b,d} and A.G. Williams\supit{a,b}
\skiplinehalf
\supit{a}Special Research Centre for the Subatomic 
Structure of Matter (CSSM),
 University of Adelaide, Adelaide, SA 5005, Australia \\
\supit{b}Department of Physics, University of Adelaide, Adelaide, SA 5005, Australia \\
\supit{c}Department of Physics, Hasanuddin University, Makassar 90245, Indonesia \\
\supit{d}Jefferson Laboratory, 12000 Jefferson Ave., Newport News, VA 23606, USA
}

\authorinfo{Marco Bartolozzi: E-mail: mbartolo@physics.adelaide.edu.au}

\begin{document} 
 \maketitle 

\begin{abstract}
In the past few years, several studies have explored the  topology of
interactions in different complex systems.  Areas of investigation
span from biology to engineering, physics and the social
sciences. Although having  different microscopic dynamics, the
results demonstrate that most systems under consideration tend to
self-organize into structures that share common features. In
particular, the networks of interaction are characterized by a power
law distribution, $P(k)\sim k^{-\alpha}$, in the number of connections
per node, $k$, over several orders of magnitude. Networks that fulfill
this propriety of scale-invariance  are referred to as ``scale-free''.
In the present work we explore the implication of scale-free
topologies in the antiferromagnetic (AF) Ising model and in a stochastic
model of opinion formation.  In the first case  we show that the
implicit disorder and frustration lead to a spin-glass phase
transition not observed for the AF Ising model on standard lattices. 
We further illustrate
that the opinion formation model  produces a coherent, turbulent-like
dynamics for a certain range of parameters. The influence, of random
or targeted exclusion of nodes is studied.  
\end{abstract}


\section{INTRODUCTION}
\label{sect:intro}  

Systems composed of many parts that interact with each other  in a
non-trivial way are often referred to as {\em complex systems}.  An
intriguing issue concerns the role played by the topological
structures in the dynamics of these systems.  Recent empirical studies
focusing on the  properties of interactions in different biological,
social and technological systems have made it possible to shed some
light  on the basic principles of structural self-organization.  A few
examples include  food webs~\cite{Williams02,Camacho02,Montoya02},
power grids and neural networks~\cite{Watts98,Amaral00},  cellular
networks~\cite{Jeong00,Jeong01}, sexual contacts~\cite{Liljeros01},
Internet routers~\cite{Faloutsos99,Pastor01,Yook02},  the World Wide
Web~\cite{Albert99,Kumar00},  actor
collaborations~\cite{Watts98,Albert99,Amaral00,Barabasi99}, the
citation networks of scientists~\cite{Redner98,Vazquez01} and the
stock market~\cite{Bonanno03,Onnela03}.  Although different in the
underlying interaction dynamics or {\em micro-physics}, most of the
empirical studies have shown a tendency of the networks to
self-organize in structures that share common features. In particular,
the number of connections, $k$,  for each element, or node, of the
network follow a  power law distribution, $P(k)\sim
k^{-\alpha}$. Networks that fulfill this property  are referred to as
{\em scale-free networks} (SFN).  In addition many of these networks
are characterized by a high clustering coefficient, $C$, in comparison
with random graphs~\cite{Bollobas85}. The clustering coefficient, $C$,
is computed as the average of local clustering,
$C_i=2y_i/(z_i(z_i-1))$, where $z_i$ is the total number of nodes
linked to the site $i$ and $y_i$  is the total number of links between
those nodes.  As a consequence both  $C_i$ and $C$ lie in the interval
[0,1].  The high level of clustering found, supports the idea that a
{\em herding} phenomenon is a common feature in social and biological
communities.  Numerical studies on SFNs have demonstrated how the
topology plays a fundamental role in infection
spreading~\cite{Pastor01} and tolerance against random and
preferential node removal~\cite{Albert00,Cohen00,Callaway00}.  A
detailed description of the progress in this emerging field of
statistical mechanics can be found in recent reviews~\cite{Albert02,Dorogovtsev02}.

In the following section we briefly describe the algorithm used for generating the 
SFN, that is the Barab$\acute{\rm a}$si-Albert model with tunable clustering.
We then study the implications of a SFN topology in two
different complex systems. The first one, Sec.~\ref{spin_glass}, in the
antiferromagnetic (AF) Ising model while the second, Sec.~\ref{op_form},
is a stochastic model for opinion formation.

\section{The Barab$\acute{\rm a}$si-Albert model with tunable clustering}
\label{sf_net}

 The Barab$\acute{\rm a}$si-Albert model~\cite{Albert99} is based on two
main assumptions: (i) linear growth and (ii) preferential attachment.
In practice the network is initialized with $m_0$ disconnected
nodes. At each step a new node with $m$ edges is added to the
pre-existing network.  The probability that an edge of the new node is
linked with the $i$th node is expressed by $\Pi(k_i)=k_i/\sum_{j}k_j$.
The iteration of this preferential growing process
yields a scale free network, $P(k)\sim k^{-\alpha}$ where the degree
distribution parameter $\alpha=3$.

It is worth noting that the Barab$\acute{\rm a}$si-Albert model  cannot
produce a high clustering coefficient. In fact, the value  of this
coefficient depends on the total number of nodes, $N$, in the
network~\cite{Albert02} and in the thermodynamic limit, $N \rightarrow
\infty$, $C  \rightarrow 0$.  In principle the observed local
clustering can play an important role in the opinion formation of
groups of people, independent of their total number. In order to
account for this, we introduce a further step in the growth process,
namely the triad formation proposed by Holme and Kim~\cite{Holme02}. In
this case, if the new added node is linked with an older node, $i$,
having other links, then with a certain probability, $\theta$, the
next link of the new node, if any remain, will be added to a randomly
selected neighbour of node $i$.  This method of introducing friends to
friends, while preserving the scale-free nature of the networks,
generates high clustering coefficients that do not depend on the
number of nodes in the network. The only tunable parameter  that
changes the value of the clustering  coefficient is the 
clustering probability $\theta$. All the simulations in the present work 
have been carried out using
$\theta=0.9$, providing to an average clustering coefficient of $C
\sim 0.39$, close to the value found in many real systems~\cite{Albert02}.

\section{Spin-glass behaviour of the Antiferromagnetic Ising model on scale-free
network}
\label{spin_glass}

The ubiquity of SFNs in nature has inspired physicists to investigate
the dynamics of standard models in the new case where the interactions
between elements are described by complex interactions.  These include
the study of various magnetic models such as the Ising model.  An
intriguing issue concerns how the unusual topology acts to influence
the cooperative behaviour of the spins.  Studies of the ferromagnetic
(FM) Ising model on a SFN, using several theoretical techniques~\cite
{Aleksiejuk02,Dorogovtsev02b,Igloi02,Herrero04} including the Monte
Carlo (MC) method~\cite{Herrero04}, have found the robustness of
ferromagnetic ordering against thermal fluctuations for the degree
distribution exponent $\alpha \leq 3$.

The robustness feature is naturally expected as SFNs have large
connectivities.  This is analogous to the FM Ising model on a regular
lattice above the lower critical spatial dimension, $d_l= 2$.  There the
ordered phase is very robust against thermal fluctuations.  However,
for the antiferromagnetic (AF) case with a SFN, the situation is
different.

Two factors come to play a central role in the dynamics of the AF-SFN
model; namely the competition induced by the AF interaction in the
elementary triangles of the network and the randomness related to the
non-regular connections.  The abundance of elementary triangles in the
network leads to frustration, as, for example, only two of the three
spins can be anti-aligned.  More generally, frustration refers to the
inability of the system to remain in a single lowest energy state
(ground state).  These ingredients lead the AF SFN to belong to a
class of randomly frustrated systems commonly referred to as spin
glasses (SGs).

Most studies of SGs have been performed on regular lattices.  These
studies have shown that frustration and randomness are the key
ingredients for SG behavior, characterized by a frozen random spin
orientation at low temperatures~\cite{Binder86}.
A study of the AF Ising model on a SFN is of great theoretical
interest since, in fact, it does possess all the characteristics of a
SG. Reviews on SG can be found in Refs.~\cite{Binder86}.
  
We consider the AF Ising model on a Barab$\acute{\rm a}$si-Albert
network with a tunable clustering coefficient, as described in
Sec.~\ref{sf_net}.  We illustrate that the AF model undergoes a SG
transition.  Such a transition is not observed on a regular triangular
lattice where, for the AF Ising model, the spins are fully frustrated.

\subsection{Model and Simulation Method}\label{two}

On each SFN constructed at the beginning of the simulation, we assign
to each vertex an Ising spin, and to each link an AF interaction.  The
Hamiltonian can be written as follows
\begin{equation}\label{ham}
  H = -\sum_{\langle ij \rangle} J_{ij}\,  s_i\, s_j \, .
\end{equation}
Here the summation is performed over the connected spins $s_i$ and
$s_j$ occupying sites $i$ and $j$, respectively. The coupling
interaction $J_{ij} = J < 0$ is AF.
As previously mentioned, each vertex with the local  cluster
coefficient $C_i > 0$ together with its neighbours, compose elementary
triangles.  Due to the AF interactions the local system is frustrated. 

It is worth pointing out that $C$ is related to the degree of
frustration of each network.  Due to the probabilistic algorithm used
for their construction, the value of $C$ fluctuates from one network to
the next.  This property is not shared by other algorithms which use
recursion formulas to generate scale-free structures, such as, for
example, the Apollonian networks~\cite{Andrade05}.

As a random system, each realization of a network of size $N$ will
differ in the ``structure'' of connectivities.  Therefore, in order to
have reliable statistics, we average over many realizations of the SF
network for each specified size.  In general, one takes into account
more realizations for small system sizes and less for large system
sizes as the latter tend to self-average.  The system sizes that we
simulate are $N =$ 1024, 2048, 4096, and 8192.  Since the
self-averaging of physical quantities for larger system sizes are
interfered by the increase of ground state degeneracy, we did not take
less realizations. Instead all physical quantities of interest for
each system size are averaged over 1000 network realizations.

Another peculiarity of SF networks regards the existence of a broad
distribution of ``hubs'', that is nodes with a large number of
connections, $k$.  The energy difference in a spin flip actually
depends on the number of connections of the spin itself, $\Delta
E_{i}= -2 s_i \sum_{j=1}^{k_i}s_j$.  Thus in the AF case for the $i$th
spin with $k_i$ connections, the hubs are more likely to ``freeze''
into a particular configuration compared to the nodes with just few
links.  This fact resembles the spin glass behaviour of particular
alloys where some elements freeze into a particular orientation at a
higher temperature than others.

The calculation of the thermal averages of the physical quantities of
interest is performed using the replica exchange MC
method~\cite{Hukushima96}, appropriate for systems such as spin-glass.
For a given network configuration, replicas having an associated
inverse temperature, $\beta$, are created.  In using this method, we
define a ``local'' MC (LMC) update as a MC update for each spin of
each replica, either consecutively through all elements of the network
or randomly.  Given that we can group the inverse temperatures in even
and odd pairs, $(\beta_{m},\beta_{m+1})$, after each LMC update we
alternate attempts to switch configurations from one temperature to
the next.  According to this procedure, we define a Monte Carlo step
(MCS) as a LMC plus a half ($m$ odd or even) exchange trial.  

For each network realization we run $3 \times 10^5$ MCSs after a
transient period of $10^3$ LMC updates.  We take a total of $6 \times
10^4$ measures for the thermal averages.  The simulation is run down
to low temperatures in a search for the possible existence of a phase
transition.  All the thermal averages obtained are then averaged over
the whole ensemble of networks.  In the following, we indicate
$\langle...\rangle$ as the thermal average and $\left[...\right]_{av}$
as the ensemble average.  The statistical errors in the plots, where
reported, are calculated via the bootstrap method.

\subsection{Observing Spin Glass Behaviour}

With the presence of frustration and randomness in the AF-SFN model,
we expect to observe a spin glass transition, i.e., a transition from
a temporal disordered to a temporal ordered phase at low temperatures.
A  quantity that is often used to characterize the SG state is the
overlap parameter, $q$,  defined as~\cite{Parisi83}
\begin{equation}\label{qorder}
q = \left[\langle \frac{1}{N}\sum_{i} s_i^{\alpha}\, s_i^{\beta} \rangle\right]_{av},
\end{equation}
where the superscripts $\alpha$ and $\beta$ denote two copies of the
same configuration of connectivity at the same temperature.

In particular, for the Ising system, due to the $Z_2$ symmetry, it is
important to evaluate the absolute value of the order parameter, $ |q|
= \left[\langle |1/N\sum_{i} s_i^{\alpha}\, s_i^{\beta}|
\rangle\right]_{av}$, to overcome the implication of the $Z_2$
symmetry of the Hamiltonian.  That is, if the system is at thermal
equilibrium and if we take quite long MCS then the usual $q$ should
average out and give an approximately zero value. The existence of a
spin glass phase is indicated by the convergence of $|q|$ to a finite
value as we increase the network size and, at the same time, a
convergence of $|q|$ to zero at high temperatures.  In the latter case
the system is in the paramagnetic phase.

\begin{figure}
\vspace{1cm}
\begin{center}
\begin{tabular}{c}
\includegraphics[height=6cm,width=14cm]{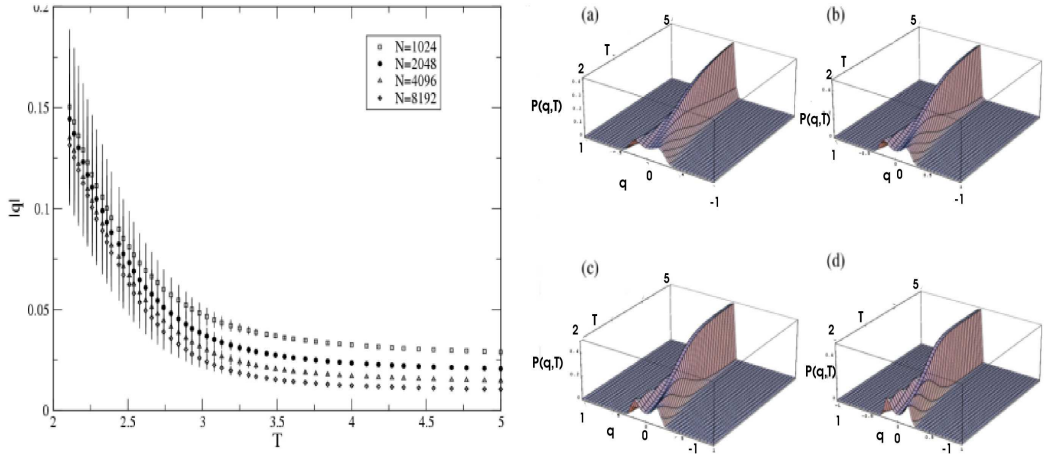}
\end{tabular}
\end{center}
\caption{Left: Temperature dependence of the overlap parameter, $q$,
for different system sizes $N$.  The increasing value of $q$ at low
temperatures indicates a SG phase.  For a given network size, 1000
realizations of the SFN are averaged over.  Right: The distribution of
$q$ at various temperatures for different system sizes, including (a)
$N=1024$, (b) $N=2048$, (c) $N=4096$ and (d) $N=8192$.  The
temperatures are provided in units of $J/k_B$, where $k_B$ is the
Boltzmann constant.}
\label{overlap}
\end{figure}

The temperature dependence of $|q|$, resulting from the simulations,
is shown in Fig.~\ref{overlap} (Left).  The existence of a SG phase is
indicated by the finite value of $|q|$ in the low temperature region,
and the approach of $|q|$ to zero at higher temperatures associated
with paramagnetic phase.  For high temperatures and large networks,
$|q|$ is approaching zero in accord with the thermodynamic limit where
$|q| = 0$.

The existence of these two different phases can also be observed from
the distributions of $q$, as shown in Fig.~\ref{overlap} (Right).  For
higher temperatures we observe simple Brownian fluctuations of the
values of $q$, leading to a singly peaked Gaussian distribution
characteristic of a paramagnetic state.  By decreasing the
temperature, the distribution starts to spread out, reflecting the
increasing number of metastable disordered states reflecting the
presence of substantial frustration.  At lower temperatures the
distribution develops double peaks associated with the Edward-Anderson
parameter representative of the SG phase.  The transition between
these two phases is roughly estimated at $T \sim 4$.  We note that the
shape of the observed distribution is different from that of the
conventional Ising system where the double peaks approach delta-like
double peaks reflecting a simple doubly degenerate ground state.

A more accurate evaluation of  the phase transition  is done through
the Binder parameter defined as follows
\begin{equation}
g_L = \frac{1}{2}\left(3 - \frac{\left[\langle
    q^4\rangle\right]_{av}}{\left[\langle q^2\rangle
    \right]_{av}^{2}}\right) \, ,
\end{equation}  
where $\langle q^2 \rangle$ and $\langle q^4 \rangle$ are respectively
the second and the fourth cumulant moment of $q$ and $0 \le g_L \le
1$.
At high temperature, when the thermal fluctuation overcomes all
cooperative interaction, the system is expected to exist in the
paramagnetic phase where there is no spatial nor temporal
autocorrelation.  As a result, the distribution of $q$ should be
Gaussian centered at $q=0$.  In this case the ratio of the cumulants,
$\langle q^4 \rangle /\langle q^2 \rangle^2 \rightarrow 3 $, resulting
in $g_L \rightarrow 0$.
At low temperatures, the cooperative interaction becomes dominant and  the
ratio of the cumulants approaches unity so that $g_L = 1$.  

\begin{figure}
\vspace{1.5cm}
\begin{center}
\begin{tabular}{c}
\includegraphics[height=6cm]{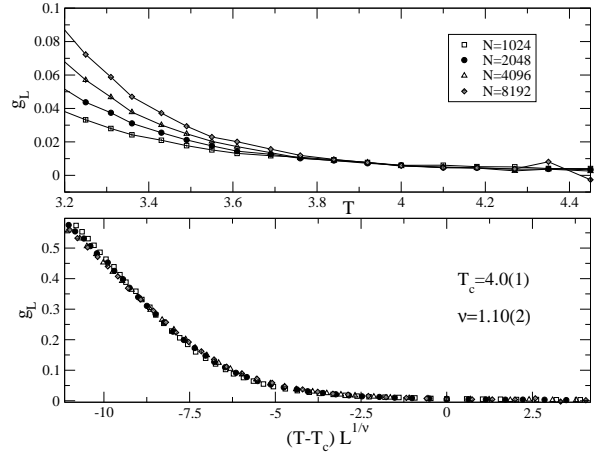}
\end{tabular}
\end{center}
\caption{Top: Scaling behaviour of the Binder cumulant, $g_L$, for
  different system sizes.  Each system size is averaged over 1000
  realizations of the network configuration.
  Bottom: Scaling plot of the data illustrated above (Top), fit to
  Eq.~\protect\ref{scalebind}.} 
\label{fig_bind}
\end{figure}

Fig.~\ref{fig_bind} (Top) displays the temperature dependence of the
Binder parameter for a variety of network sizes.  A spin glass state
is observed for lower temperatures where the Binder parameter deviates
from zero, and increases with the system size.  In the thermodynamic
limit, we expect $g_L \to 1$ just below the critical temperature.  A
crossing point in the size dependence of $g_L$ indicates that the
critical temperature for the SG phase transition is $T \sim 4.0$.
For temperatures above $T \sim 4.0$ the Binder parameter, while
remaining always above zero, does indeed order in an opposite manner
indicative of a genuine crossing of the curves and in accord with a
genuine spin glass transition at finite temperature.

A more reliable estimate of the critical temperature, $T_c$, for
finite size systems can be given by using scaling arguments.  For a SG
system, the Binder parameter depends on the system size
$L$ as
\begin{equation}\label{scalebind}  
g_L = \tilde g_L \, [(T-T_c) \, L^{1/\nu}] \, ,
\end{equation}
with $\nu > 0$.
At $T_c$ the Binder cumulant does not depend on $L$.
For the SFN, the system size scales logarithmically with the number of
nodes $N$~\cite{Albert02}, and therefore we take $L = \log(N)$.
The parameters $T_c$ and $\nu$ are determined by constraining the
temperature dependence of the Binder parameter for each network size
to lie on a single curve.  The curves following the scaling bahaviour
of Eq.~(\ref{scalebind}) are shown in Fig.~\ref{fig_bind} (Bottom).
>From this fit we estimate the critical temperature $T_c\sim 4.0(1)$
and the exponent of the SG correlation length $\nu \sim 1.10(2)$.
It is important to underline that this kind of behaviour is not observed
for an AF system on a regular triangular lattice.

\section{Stochastic model of opinion formation on a scale-free network}
\label{op_form}

We turn now our attention to the role played by the SFN on a model
of stochastic opinion formation. In this case,
once the scale-free network has been built, we randomly assign   the
spin values, $\pm 1$, to  every node. These values 
correspond to a Boolean kind of opinion while the bonds of
the networks represent the interactions between agents.

The dynamics of the spins follows a stochastic process
that mimics the human uncertainty in decision
making~\cite{Krawiecki02,Bartolozzi04}.  Values are
updated synchronously according to a local probabilistic rule:
$\sigma_{i}(t+1)=+1$ with probability $p_{i}$ and $\sigma_{i}(t+1)=-1$
with probability $1-p_{i}$.  The probability $p_{i}$ is determined, by
analogy with heat bath dynamics with formal temperature $k_{B}T=1$,
$p_{i}(t)=1/(1+e^{-2I_{i}(t)})$, where the local field, $I_{i}(t)$, is
\begin{equation}
I_{i}(t)=a \xi(t)\tilde{N_i}^{-1}\sum_{j=1}^{\tilde{N_i}}\sigma_{j}(t)
+ h_i \eta_{i}(t) r(t).
\label{field}
\end{equation}
The first term on the right-hand side of Eq. (\ref{field}) represents
the  time dependent interaction strengths between  the node $i$ and
his/her $\tilde{N_i}$  information sources, which are the first
neighbours in the network.  The  second term  instead reflects the
personal reaction to the system feedback, that is the average opinion,
$r(t)=1/N\sum_{j=1}^N \sigma_j(t)$, resulting from the previous time step.  
The terms $\xi(t)$ and
$\eta_{i}(t)$ are random variables uniformly  distributed in the
interval (-1,1) with no correlation in time nor in the network. They
represent the conviction, at time $t$,
 with which agent $i$ responds to his/her group (common for all the agents)
 and the global opinion of the network respectively.
The strength term,  $a$, is  constant and common for the whole network,
while  $h_i$ is specifically chosen for every individual from a
uniform distribution in (0,$\kappa$) and are both constant  in the
dynamics of the system. By varying the parameter $\kappa$ we can give
more or less weight to the role of feedback in the model.  The
strength coefficients $a$ and $h_i$ in the local field, $I_i$,
characterizing the attributes of the agents, play a key role in the
dynamics of the model.  They represent the relative importance that
each agent of the network gives, respectively, to his/her group and
to the variation of the average opinion itself.
While $a$ is a parameter associated with the network, $h_i$ is specifically
chosen for each individual at the beginning of each simulation.

\subsection{Numerical Simulations}
\label{numerical}

\begin{figure}
\begin{center}
\begin{tabular}{c}
\includegraphics[height=6cm]{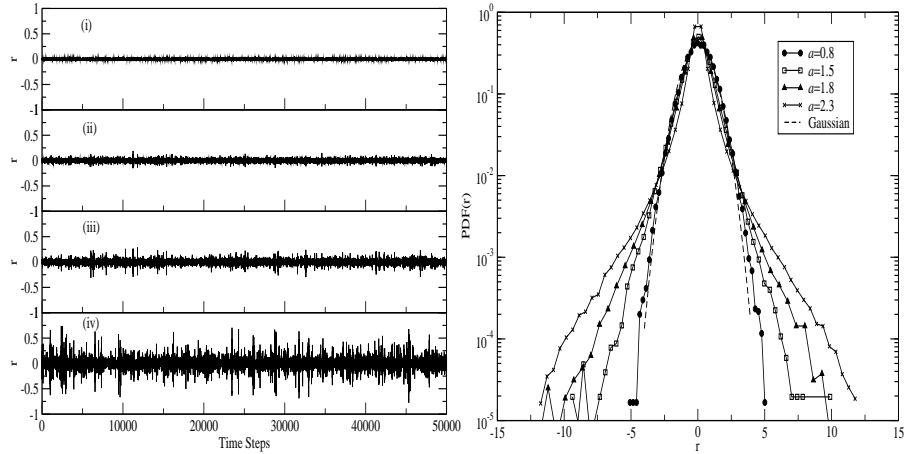}
\end{tabular}
\end{center}
\caption{\label{ts_a} Left: Time series of the average opinion, $r$,  for different
values of the group interaction strength parameter
 $a$: (i) $a=0.8$, (ii) $a=1.5$, (iii) $a=1.8$
and (iv) $a=2.3$. The parameters used for the simulations are: $N=10^4$
nodes, clustering probability  $\theta=0.9$, initial nodes and links 
per new node $m_0=m=5$
 and we take the upper bound of the distribution of personal response strengths equal
to the group interaction strength, that is $\kappa=a$. The results
involve 10 realizations of the scale free network each displayed for
5000 time steps.  For values of $a$ greater than 1 a turbulent-like
state, characterized by large fluctuations, starts to appear in the
process of opinion formation.  Right: 
PDFs of the time series relative to the time series in (Left). The shapes of
the distributions converge to a Gaussian for small values of the group interaction 
strength $a=\kappa$. A Gaussian 
distribution is also plotted for comparison. All the PDFs in this paper are
obtained over 50 realizations of the SF network.
In order to compare the fluctuations at different scales, 
the time series in the plot have been normalized according to 
$r(t) \rightarrow \frac{r(t)-\bar{r}}{\sigma}$, where $\bar{r}$ and $\sigma$ 
denote the average and the standard deviation over the period considered respectively. }
\end{figure}

At first we investigate the importance of the group strength $a$ 
by fixing $\kappa=a$.  
In this case the dynamical behaviour is similar to that found 
in the stock market context in Refs.~\cite{Kaizoji00,Krawiecki02,Bartolozzi04}. 
For $a \tilde{<} 1$  the resulting time series of average opinion is largely
uncorrelated Gaussian noise with no particularly 
interesting features, as illustrated in Fig.~\ref{ts_a}(i) (Left).

As soon as we exceed the value of $a \approx 1$ a turbulent-like
 regime sets in, characterized by large intermittent fluctuations,
 Fig.~\ref{ts_a}(ii $\rightarrow$ iv) (Left).  These large
 fluctuations, or   {\em coherent events}, can be interpreted in terms
 of a multiplicative stochastic process with a weak additive noise
 background~\cite{Nakao98,Krawiecki02}.  For $a > 2.7 $ we observe
 that the bursts of the time series begin to  saturate the bounds $-1
 \le r \le 1$.

In Fig.~\ref{ts_a} (Right) we plot the probability distribution functions (PDFs) 
associated with the time series of Fig.~\ref{ts_a} (Left). The large fluctuations, 
for $a$ greater than $\approx 1$, are reflected in the fat tails of the relative PDFs.
Decreasing the value of $a$, and so the number of coherent events,
 the PDF converges to a Gaussian distribution generated by a
 random Poisson process.

In order to test the relevance of the network structure on the process
of opinion formation, the previous simulations have been repeated,
with a large number of nodes, $N$, and $\kappa=a$,  for different
values of the clustering parameter, $\theta$, and the node-edge
parameter, $m$.  While varying $\theta$,  does not lead to any
substantial difference in  the dynamics of the model,  the increase of
the average number of links per node, $\bar{k}=2m$, has a dramatic
effect in the turbulent-like phase, which deviations from a Gaussian regime 
increase dramatically: large scale synchronizations
are more likely to occur for large $m$. 
 This behaviour is
intrinsically related to the model of Eq.~(\ref{field}). In fact, the
turbulent-like regime is a consequence of the random fluctuations of
the interaction strengths between agents  around a  bifurcation value
separating the ordered and disordered phase.

It is also worth pointing out that an increase of $\bar{k}$ is 
related to a decrease in the average path length between nodes; that is, the
network ``shrinks'' and becomes more compact. In relation to our
previous discussion, the more compact the network is the more the
dynamics of our system approaches to the mean field approximation.
It becomes easier for the agents to synchronize.  This characteristic of
compactness, referred to as the {\em small world
effect}~\cite{Bollobas85,Albert02,Dorogovtsev02}, is actually very common in both
real and artificial networks.

These results confirm that the critical topological characteristic
leading to herding behaviour in the framework of stochastic  opinion
formation is the presence of mean field effects enhanced by
small-world structure.  The more information (links) that an agent
has, the more likely it is for him/her to have an opinion  in accord
with other agents.

\subsection{The Influence of Indecision}
\label{indecision}

We now extend our model in order to include the concept of indecision.
 In practice a certain agent $i$, at a time step $t$, may take neither
 of the two possible decisions, $\sigma_i= \pm 1$, but remain in a
 neutral state.  Keeping faith to the spirit of the model, we address
 this problem introducing an {\em indecision probability}, $\epsilon$:
 that is the probability to find, at each time step, a certain agent
 undecided.  This is equivalent to introducing time dependent failures
 in the structure of the network by setting $\sigma= 0$.

\begin{figure}
\begin{center}
\begin{tabular}{c}
\includegraphics[height=6cm]{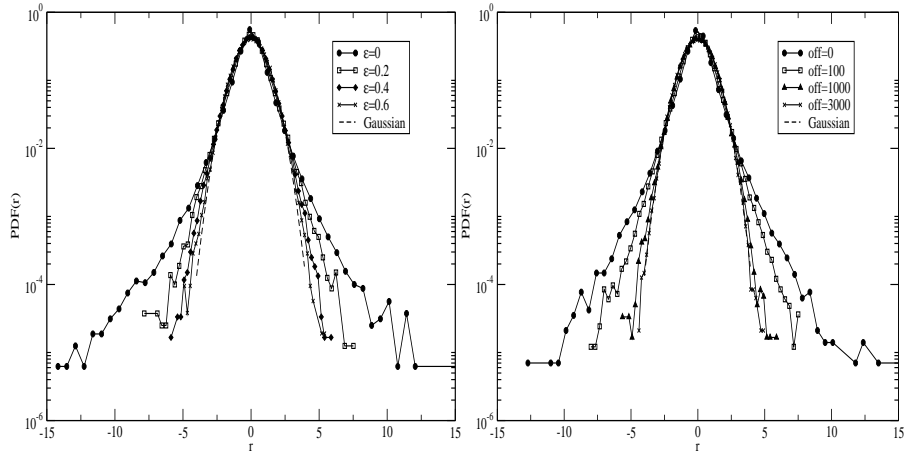}
\end{tabular}
\end{center}
\caption{ \label{pdf_rand} Left: Transition from coherent bahaviour,
 indecision probability $\epsilon=0$, to noise
using a random selection for the inactive agents. For $\epsilon \approx 0.6$ we
reach a noise-like behaviour. The parameters used in the simulation are:
$N=10^4$ nodes, $\theta=0.9$ for the clustering probability, $m=m_0=5$ for the
links of each new node, $a=1.8$ and $\kappa=a$ for the group and global opinion response
respectively. Right: In this simulation we progressively turn off the largest 
hubs in the network. Once we have turned off about the 10\% of agents, $N=10^4$,
the coherence in opinion formation disappears. }
\end{figure}

Focusing on the turbulent-like regime, the shape of the PDF in the
opinion fluctuations changes according to different concentrations of
undecided persons.  The results of the simulations,  in
Fig.~\ref{pdf_rand} (Left), show how the dynamics of the model move from an
intermittent state for $\epsilon=0$ toward a noise state for $\epsilon
\approx 0.6$.  The convergence to a Gaussian distribution is obtained
only for quite  high concentrations of undecided agents at about 60\%.
The robustness of the turbulent-like behaviour is related to the
intrinsic robustness of SF networks against random
failures~\cite{Albert00,Cohen00,Callaway00}.  In fact, because there is a large absolute
number of poorly connected nodes, related to the power law shape of
$P(k)$, the probability of setting one of them to inactive is much
higher compared to the ``hubs'' that are relatively rare.

We can claim that, in large social networks governed by stochastic reactions
in their elements, large fluctuations
in the average opinion can appear even in the case in which a large part of
the network is actually ``inactive'' provided that the structure is scale
free and the indecision is randomly distributed.
The existence of large hubs provides for the survival of extended sub-networks
in which synchronization can give rise to coherent events. The structure
of the network itself supplies the random indecision.

Now we address the question of how the dynamics may change if we do
not choose randomly the inactive nodes but we target the nodes having
the most links.  What we do in practice is to sort the nodes according
to their number of links and then deactivate the nodes having the largest
number of links in decreasing order.  Fig.~\ref{pdf_rand} (Right) illustrates
how the fragmentation process is much faster and the noise regime is
reached already when only the 10\% of the hubs are deactivated.  As
emphasized in Ref.~\cite{Albert00,Cohen00,Callaway00}, the hubs have a great importance
in the structural properties of SF networks and specifically targeting
these nodes can lead to sudden isolation of a large fraction of the
nodes of the network.

\section{Discussions and Conclusions}

Motivated by recent empirical findings for complex systems interaction
topologies, we have explored the implications of a Barab$\acute{ \rm
a}$si-Albert SFN topology in two different models of complex systems.
In particular we have found that the random frustration introduced by
this topology of interactions induces a transition from a paramagnetic
state to a spin glass state for an AF Ising model at a finite
temperature.  The critical spin glass phase transition temperature is
estimated to be $T_c \sim 4.0(1)$.  Such behaviour is not observed for
the AF Ising model on regular lattices.

The SFN topology also has important consequences in our model for
opinion formation.  In this case, we discovered that the ``hubs'' of
the network are more likely to synchronize due to mean field effects.
Conversely, these effects are not strong enough to synchronize the
poorly connected nodes.  Moreover, introducing inactive agents and
spreading them randomly on the network, does not spoil the
turbulent-like state, even for high concentrations of ``gaps'' up to
approximately 60\% of agents.  This is a consequence of the implicit
robustness of SF networks against random failures.  If instead of
selecting randomly the undecided individuals, we aim directly to the
``hubs'' of the network then the situation changes.  In this case the
network is disaggregate, composed of very small sub-networks and
isolated nodes. Synchronization cannot significantly effect the
resulting global opinion and the time series approximates Gaussian
noise.

\acknowledgments     
 
Part of  the computation of this
work has been done using super computer facilities of the South
Australian Partnership for Advanced Computing (SAPAC).


\end{document}